\documentclass[twocolumn,prd,showpacs,preprintnumbers,amsmath,amssymb,amsfonts,nofootinbib]{revtex4}

\usepackage{graphicx}
\usepackage{amssymb}
\usepackage{psfrag}
\usepackage{ulem}
\usepackage{color}
\usepackage{hyperref}
\usepackage{natbib}
\definecolor{dr}{rgb}{1.00,0.0,.33}
\begin{document}

\title{Viability of the cluster mass function formalism in parametrised modified gravity}

\author{Daniel B.~Thomas}
\email{daniel.b.thomas08@imperial.ac.uk}
\author{Carlo R.~Contaldi}
\affiliation{Theoretical Physics, Blackett Laboratory, Imperial
  College London, London SW7 2BZ, UK}
\date{\today}


\begin{abstract}
  Model-independent parametrisations for examining departures from
  General Relativity have been increasingly studied over the past few
  years. Various observables have been used to constrain the
  parameters and forecasts for future surveys have been carried
  out. In one such forecast \cite{paper2}, galaxy cluster counts were
  used to constrain the parameters. Here, we carry out a limited set
  of $N$-body simulations, with a modified Poisson equation, to
  examine the accuracy of existing mass functions for modified gravity
  cosmologies. As well as altering the gravitational calculation, we
  include the effect of a screening scale to ensure consistency of the
  theory with solar system tests. Our results suggest that if a
  screening scale exists its effect can be taken into account in the
  cluster count calculation through its effect on the linear matter
  power spectrum. If this is done, the accuracy of the standard mass
  function formalism in modified gravity theories with reasonably
  small departures from General Relativity, as tested in this work, is
  comparable to the standard case.
\end{abstract}

\maketitle

\section{Introduction}

Several explanations have been put forward to explain the apparent
accelerating cosmological expansion. The most popular idea is that the
acceleration is due to a distinct form of energy in the universe,
generically known as dark energy. The simplest example of what could
constitute dark energy is a cosmological constant $\Lambda$, a
non-dynamical form of energy that may be interpreted as arising as an
unconstrained integration constant in the equations of motion or from
the vacuum energy generated by the quantised fields that make up the
matter content of the universe. Dynamical forms of dark energy
involving scalar fields that are evolving with time can also lead to
late-time acceleration. Some have also argued that the observation
that the expansion is accelerating results from the misuse of the
Cosmological Principle in interpreting the expansion history from the
inhomogeneous distribution of matter around us. An alternative
perspective to the above both lines of thought above is to explain the
departure from the expected late-time behaviour by introducing
modifications to the theory of General Relativity (GR) itself.

There has been an increased interest in modified gravity theories in
cosmology over the past decade or so. The large number of proposed
alternatives to General Relativity (GR) has led to interest in
examining possible deviations from GR in a more model independent
way. Much work has gone into creating these parametrisations,
examining their consistency, constraining them with current data and
forecasting future experiments (see \cite{mgreview} for a
comprehensive review). Many different observations and combinations of
observations have been considered for constraining these
parametrisations. The key point with the combinations is the issue of
distinct degeneracies in different sets of observations. In
particular, observations such as the integrated Sachs-Wolfe effect in
the Cosmic Microwave Background (CMB) and cosmic shear surveys, both
being effects on the energies and trajectories of photons that are
relativistic, constrain the sum of the two scalar potentials in the
metric $\Psi + \Phi$ and thus constrain a particular combination of
the Modified Gravity Parameters (MGPs) as will be introduced
later. This degeneracy is best broken with an observation that depends
on a different linear combination of the scalar
potentials. Observations relating to the growth of structure in
non-relativistic matter, such as redshift space distortions or galaxy
surveys, are a natural choice for breaking the degeneracy. Recently
\cite{paper2}, we argued that combining future cluster counts with CMB and
cosmic shear observations could yield high precision constraints on
the simplest form of MGPs particularly in the case where the
modifications are constant at low redshift.

Galaxy cluster counts can be predicted for a particular combination of
cosmological parameters by starting from the linear matter power
spectrum and using mass functions to calculate how many bound
structure will form at different mass scales. The mass functions are
semi analytic formulas whose form derives from the original
Press-Schechter ideas \cite{ps} and its extensions
\cite{bond91,st99a,st99b,st02}. The functions are now calibrated using
detailed and extensive $N$-body simulations. In this \textit{paper},
we present the results of a limited set of of $N$-body simulations
aimed at investigating the accuracy of existing mass functions for the
MGP formalism. Due to the computational demand of running the required
$N$-body simulations we have restricted our investigation to a single
point in the phase space of MGPs i.e. a single choice in the
modification of the strength of gravity that is constant at low
redshift. As such we are not attempting to accurately calibrate an
extension of the mass function formalism in the phase space of
MGPs. We are instead addressing the question of whether the existing
mass function formalism can be used {\sl as is} in forecasts involving
cluster counts such as that carried out in \cite{paper2}.

This {\sl paper} is organised as follows. In section~\ref{sec:massf} we
discuss the mass function formalism and some of the results in the
literature regarding modified gravity models. The modifications
required to run $N$-body simulation in such models of parametrised
modified gravity are summarised in section~\ref{sec:nbody} and the
results of running our suite of simulations are described in
section~\ref{sec:res}. We conclude with a discussion of the results in
section~\ref{sec:disc}.

\section{Mass Functions}\label{sec:massf}

Galaxy clusters are some of the largest collapsed structures in the
universe. According to the standard scenario of hierarchical structure
formation within the $\Lambda$CDM cosmological model, clusters
typically consist of hot gas bound in a large cold dark matter
halo. Clusters are a useful cosmological probe as their size
corresponds to scales near the linear to non--linear transition in the
underlying dark matter power spectrum. This has several consequences:
they probe the tail of the matter perturbation spectrum and are
therefore a sensitive probe of growth. In addition, galaxy cluster
counts can be predicted from linear theory, using semi-analytic
formulae or formulae calibrated from $N$-body simulations. Cluster
counts and the properties of clusters have been used to constrain
cosmological parameters and in particular they have been used to
constrain dark energy, which may be responsible for late--time acceleration
of the cosmological expansion (see for example
\cite{wang98,battye03}), and some studies looking at constraining the
growth factor $\gamma$ (see for example \cite{rapetti10,shapiro10}).

The number of clusters observable over a fraction of the
sky $f_{\rm sky}$ and with a redshift dependent mass resolution limit
$M_{\rm lim}(z)$ in a redshift bin spanning the interval $z$ to $z +
\Delta z$ can be calculated by integrating the comoving number density
$dn/dM$ of objects with mass $M$
\begin{equation}
N_{\Delta z}=4\pi f_{\rm sky}\int^{z + \Delta z}_{z}dz'
\frac{dV}{dz'\,d\Omega}\int^{\infty}_{M_{\rm lim}(z')}\frac{dn}{dM}\,dM\,,
\label{clusterbins}
\end{equation} 
where $dV/dzd\Omega=r^2(z)/H(z)$ is the comoving volume at redshift
$z$ in a flat universe, with $H(z)$ the Hubble rate and
$r(z)=\int^{z}_{0}dz'/H(z')$ the comoving distance to that
redshift.

The comoving number density of objects in a given mass range $dn/dM$
is known as the mass function and much work has gone into predicting
its shape for a given linear power spectrum. Early, semi-analytical
estimates of the mass function resulted in the Press-Schechter
formalism \cite{ps}, where the mass function is given by
\begin{equation}
  \frac{dn}{dM}=-\sqrt{\frac{2}{\pi}}\frac{\rho}{M}\frac{d\sigma_M}{dM}\frac{\delta_c}{\sigma_M^2}\exp\left[-\frac{1}{2}\frac{\delta^2_c}{\sigma_M^2}\right]\,,
\end{equation}
where $\rho$ is the background density of dark matter today,
$\delta_c=1.686$ is the critical density contrast for collapse of a
spherical perturbation and $\sigma_M^2$ is the variance of the dark
matter fluctuations in spheres of radius $R=(3M/4\pi\rho)^{1/3}$
defined by the integral of the linear matter power spectrum $P(k)$
over wavenumber $k$
\begin{equation}
\sigma^2_M=\frac{1}{2\pi^2}\int^{\infty}_{0} W^2(kR) \,P(k) \,k^2\,dk\,,
\end{equation}
with top-hat filter function
\begin{equation}
W(kR)=3\left(\frac{\sin(kR)}{(kR)^3}-\frac{\cos(kR)}{(kR)^2} \right)\,.
\end{equation}

Successive studies have refined the formalism resulting in more
complex expressions with increased accuracy
\cite{bond91,st99a,st99b,st02}. The functional forms of these mass
functions have been used to fit the results of $N$-body simulations for
$\Lambda$CDM cosmologies and agreement is at the 5-10\% level for
$\Lambda$CDM cosmologies\cite{massf10,courtin10}. The inaccuracy
increases to 10\% or more for more general cosmologies
\cite{massf10,courtin10}. In the what follows we select the following
form from \cite{warren06}, henceforth W05:
\begin{eqnarray}\label{eq:mass}
  \frac{dn}{dM}&=&-0.7234\frac{\rho}{M}\frac{d\sigma_M}{dM}\frac{1}{\sigma_M}\exp\left[\frac{-1.1982}{\sigma^2_M}\right]
  \times \nonumber\\
  &&\left(0.2538+\sigma^{-1.1625}_M \right)\,.
\end{eqnarray}

The mass function formalism has been used extensively in constraining
extensions to the standard $\Lambda$CDM cosmology.  Some examples are
redshift dependent dark energy equation of state models $w$CDM
\cite{massf10,courtin10} and some models of modified gravity including
$f(R)$ \cite{schmidt09c,schmidtfr}, DGP \cite{schmidt09b,chan09} and
coupled scalar field cosmologies \cite{zhao10,li2010}.  However, in
some cases it is unclear how well the mass function formalism will
reproduce the correct cluster counts if the modifications alter
significantly the process of structure formation. 

There are three reasons why the expressions commonly used to predict
cluster counts may not be valid in modified models. Firstly the linear
matter power spectrum the formalism is based on will, in general,
differ from the standard cosmology for the same matter
content. Secondly the expressions are calibrated with respect to
standard $\Lambda$CDM $N$-body simulations or semi--analytical arguments
and the coefficients obtained in this way may not be the correct one
in modified cosmologies. As such, even if two models may give the same
linear matter power spectrum, they could differ in their cluster
counts predictions because, for example, the critical density
$\delta_c$ may be different in the modified model. Thirdly, the
expressions themselves may not be a corrected parametrisation of the
transformation of a linear power spectrum into a description of the
mass distribution of clusters.

Typically only the first possibility is addressed in works where cluster
counts are used to constrain modified cosmologies. This is achieved by
inputing into the formalism a modified linear matter power spectrum
that has been obtained by solving the linear growth equations for the
dark matter. This is by far the simplest way to account for the
modifications since the second and third concerns require the use of
$N$-body runs to be addressed.

Here  we will focus on how well the mass formalism does in
reproducing parametrised modified gravity cluster counts if the
expressions are unchanged and the modifications are solely taken into
account by modifications of the linear matter power spectrum. The
results will inform us on whether using the standard formalism can be
an informative tool in investigating the constraining power of cluster
counts in modified gravity models. In particular we will be able to
determine whether forecasts made in \cite{paper2} may be biased
because of the way the mass function formalism was used in the
prediction of cluster counts in modified gravity models.

\section{N-Body Simulations and Modified Gravity Parametrisation}\label{sec:nbody}

\begin{table*}[t]
\centering
\begin{tabular}{|c|c|c|c|c|c|c|c|c|}
\hline 
Gravity & $z_i$ & \#runs & $L_{\rm{box}}$($h^{-1}$Mpc) & $N_{\rm{part}}$ & $\epsilon$ ($h^{-1}$kpc) &$\mu$ &$z_{\rm{mg}}$&$\ell_{\rm{Sc's}}h^{-1}$(Mpc)\\ 
[0.5ex]
\hline
 GR  & 50 & 10 & 400 & 256$^3$ & 45 & 1.0 & n/a & n/a \\
 MGPs& 50 & 10 & 400 & 256$^3$ & 45 & 1.05 & 50 & 1.0 \\
 MGPs& 50 & 10 & 400 & 256$^3$ & 45 & 1.05 & 50 & n/a \\[1ex]
\hline
\end{tabular}
\caption{Parameters for simulations.}
\label{table:sims}
\end{table*}

Numerous sets of phenomenological ''Modified Gravity
Parameters'' (MGPs) have been suggested in the literature, see
e.g. \cite{daniel10} for a partial translation table and \cite{gbz10b}
for a discussion of the differences with some of the parametrisations.
Most of the parametrisations are phenomenological modifications to
the Einstein equations and typically involve a parameter relating to
the strength of gravity and a parameter relating the two scalar
potentials in the metric.  Our potentials are defined as scalar
perturbations of a flat, FRW metric
\begin{eqnarray}\label{eq:pert}
g_{00}&=&-\left[1+2\Psi(\vec x , t)\right]\,,\nonumber \\
g_{ij}&=&a^2(t)\left[1-2\Phi(\vec x ,t)\right]\delta_{ij}\,,
\end{eqnarray}
where we have made the conformal Newtonian gauge choice to fix the
remaining two scalar degrees of freedom in the perturbed metric.  In
(\ref{eq:pert}), $\Psi$ is the Newtonian potential and is responsible
for the acceleration of massive particles whereas $\Phi$ is the
curvature potential, which also contributes to the acceleration of
relativistic particles.  For this work, since $N$-body simulations are
essentially Newtonian plus an expanding background, we have only used
one MGP, $\mu$. This is the parameter in the Poisson equation that
controls the strength of gravity. In fourier space,
\begin{equation}\label{eq:mgp}
k^2 \Psi(a,k)=-4\pi G a^2 \mu(a,k) \rho (a)\Delta(a,k) \,.
\end{equation}
where, $a$ is the FRW scale factor, $G$ is Newton's constant, $\rho$
is the background density of cold dark matter and $\Delta$ is the
gauge invariant density contrast given by 
\begin{equation}
\Delta=\delta+\frac{3aHv}{k}\,,
\end{equation}
where the cold dark matter density contrast is defined as $\delta =
\delta\rho(a,k)/\rho(a)$, $v$ is velocity of the dark matter and $H$
is the Hubble parameter.

In principle, $\mu$ could be any function of time and space, however
here we have kept a simple treatment of it. In particular, we are
considering the situation where, at a set redshift of $z_{\rm{mg}}$,
the parameter transitions from its GR value of 1 to some other
value. Except for the inclusion of a screening scale, there is no
scale dependence inserted into the phenomenology. Screening mechanisms
are an important component of modified gravity models as they allow
models that differ from GR to reproduce solar system tests. There are
several mechanisms in the literature, including the chameleon
mechanism \cite{chameleon} that operates in $f(R)$ gravity and the
Vainshtein mechanism \cite{vainshtein} that operates in DGP
gravity. To further our phenomenological approach to modified gravity,
we use a simpler treatment where the modifications to gravity cease
below a set length scale, $\ell_{\rm{sc}}$.  

In order to investigate the use of mass functions when forecasting
cluster counts in models of parametrised modified gravity, we have
made some simple modifications to the publicly available $N$-body,
Smooth Particle Hydrodynamics (SPH) code Gadget-2
\cite{gadget2}. Gadget-2 is a parallel $N$-body code that calculates
the gravitational acceleration with a TreePM approach
\cite{treepm}. For this investigation we are only interested in the
dynamics of the dominant dark matter and we do not include any
component with non-zero pressure. Since the $N$-body simulation probes
the low-velocity, Newtonian regime on sub-horizon scales, the only
modification of gravity that affects the simulations is any
modification to the Poisson equation that encompasses all
gravitational aspects of the interaction in such codes. In Gadget-2
the Poisson equation is solved differently on large and small
scales. On large scales, a regular grid (the ``mesh'') is placed in
the simulation volume. The mass of each particle is then interpolated
onto the mesh. This discrete density field is then Fourier transformed
and the Poisson equation is solved in Fourier space. This yields the
gravitational acceleration of particles in the simulation due to
distant mass distribution in the simulation volume. On smaller scales,
the force on each particle is obtained by calculating the Newtonian
potential directly. A hierarchy of increasingly higher resolution
cells (the ``tree'') is constructed. Far from the point where the
potential is being calculated, it is sufficient to use cells of coarse
resolution containing multiple particles treated together at their
centre of mass. Nearer to the point where the potential is being
calculated, increased resolution is required until, at sufficiently
short distances, the effect of each particle is included
separately. The two components of the potential are combined via a
matching filter to obtain the potential governing the overall force
acting on each particle at every time step. Thus modifications to
gravity must be encoded in both contributions to the total potential
i.e. in both Fourier domain solution and direct force calculation.

Our modifications to Gadget-2 are as follows: Both the potential
calculated by the particle mesh and the force calculated by the tree
structure are modified by a factor of $\mu$ for distances above a
screening scale of proper length $\ell_{\rm{sc}}$. This screening
scale can be switched off such that the only modification acting is
the change in the strength of gravity. We ran the modified gravity
simulations both with and without the screening scale for a series of
simulation boxes each with the same volume and mass resolution (see
table \ref{table:sims} for the parameter values used in the
simulations). The expansion history in the simulations is that of a
standard $\Lambda$CDM cosmology for both GR and MGP runs. This is in
order to isolate the effect of the MGP on the growth of structure from
any change due to the expansion history. Each simulation is started at
a redshift $z_i=50$ and average the result of each simulation over a
number of runs with independent random seeds to reduce the effect of
sample variance in our final cluster counts.

The linear matter power spectra today used for the mass function
prediction for both the GR+$\Lambda$CDM runs and the modified gravity
runs were computed using the modified version of CAMB \cite{camb}
MGCAMB \cite{mgcamb}. We have modified MGCAMB to incorporate the same
modifications to gravity as the $N$-body simulation, namely a constant
value for $\mu$, different to the fiducial GR value of 1, that
switches on after redshift $z_{\rm{mg}}$ and with the option to
recover GR below a screening scale $\ell_{\rm{sc}}$. For all
simulations in this work we have chosen a fiducial value of $\mu=1.05$
for the modification. The second MGP, $\eta=\Phi/\Psi$, has been kept
at its GR value in the MGCAMB code and does not come in to the $N$-body
simulations. We consider a flat cosmology with the following
parameters. The dimensionless Hubble rate in units of 100 Km s$^{-1}$
Mpc$^{-1}$, $h=0.71$, the density of matter (baryons + dark matter)
and dark energy in units of the critical energy density,
$\Omega_m=0.265$, and $\Omega_{\Lambda}=0.735$ respectively, the
optical depth to recombination $\tau=0.088$, the amplitude of
primordial, super horizon curvature perturbations $A_s=2.157\times
10^{-9}$ at $k=0.05\,h$ Mpc$^{-1}$ and their spectral index
$n_s=0.963$. These parameters correspond to the WMAP 7--year best--fit
parameters \cite{wmap}. This model yields a large scale structure
normalisation of $\sigma_8=0.804$ for the standard deviations of
fluctuations on scales of $8\,h^{-1}$ Mpc.

\begin{figure}[t]
\begin{center}
\includegraphics[trim = 0mm 50mm 0mm 30mm, clip, width=3.2in]{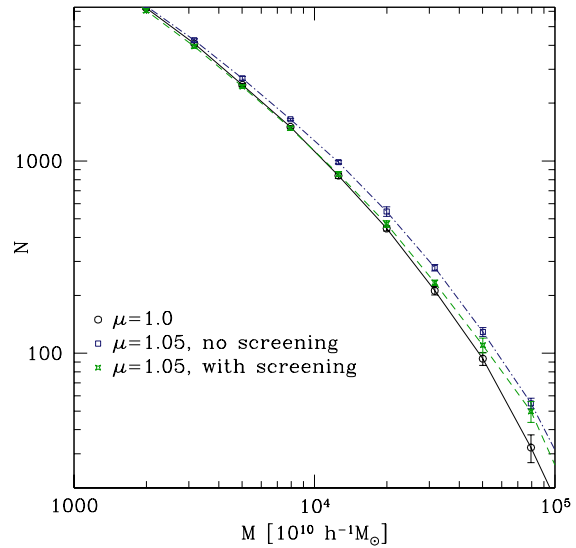}
\end{center}
\caption{The number of clusters as a function of mass of the cluster
  found in the our simulations for the GR case ($\mu=1$) compared to
  the modified gravity case, with and without screening scale. In the
  modified gravity case the 5\% increase in $\mu$ leads to an increase
  in clusters on all scales when a screening scale is not
  included. The result is more complicated when a screening scale is
  included with less cluster being produced at the low mass scales as
  expected.}
\label{fig:total}
\end{figure}

The initial condition for each of the $N$-body simulations were
calculated using the transfer functions obtained from the MGCAMB
run. The transfer functions are used to generate a grid of particles
with velocities and displacements consistent with the power spectrum
of the matter expected at the starting redshift $z_{\rm{i}}$. For this
work we have used the 2LPT \cite{2lpta,2lptb} package to obtain the
initial velocities and displacements. The package uses a second order
Lagrange perturbation scheme to evolve the displacements and
velocities of particles to the desired redshift. This is more
accurate than using the first order Zeldovich approximation.

We obtained independent initial conditions for each of the several
realisations run for each box and particle number. In the simulations
used here, all of the particles are cold dark matter particles. For
each box size and particle number, we also specified the smoothing
scale $\epsilon$.  The simulations all had a starting redshift of
$z_i=50$, however, we ran several simulations starting at higher
redshifts to check there was no effect on the final results. We have
also verified that our $\Lambda$CDM simulations reproduced the results
of earlier papers, in particular the results from
\cite{jenkins01,heit06}.

\begin{figure}[t]
\begin{center}
\includegraphics[trim = 0mm 50mm 0mm 30mm, clip, width=3.2in]{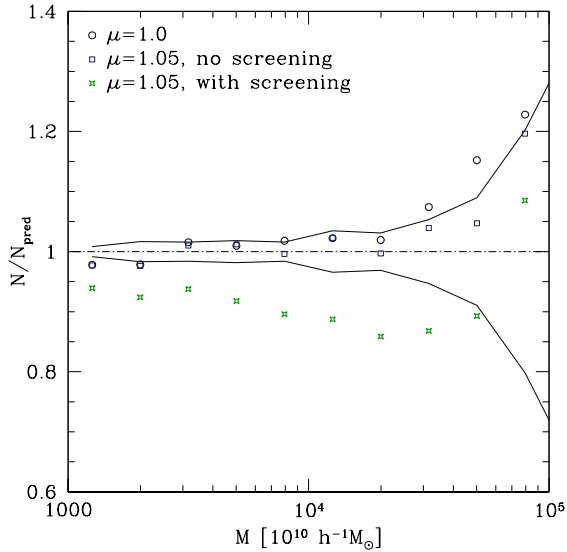}
\end{center}
\caption{The cluster count found in the GR compared to modified
  gravity and modified gravity with a screening scale of 1 Mpc for the
  simulation volume. We have divided out the respective prediction for
  GR and modified gravity based on the mass function formalism applied
  to the respective linear power spectra. The input spectrum is the
  same for both modified gravity cases in this figure. The fact that
  the prediction does not take into account the screening scale leads
  to an over-prediction of clusters on all scales. The solid line
  gives an indication of the expected sample scatter for the
  simulation based on the GR counts and is included as a rough guide
  to the significance of any discrepancy.}
\label{fig:massfunc}
\end{figure}

Once the simulations are run the challenge is to identify clusters and
their masses in order to determine the mass function of clusters for
the simulated volume.  There are two main algorithms to find dark
matter halos in cosmological $N$-body simulations, namely spherical
overdensity algorithms \cite{lacey1994} and friends-of-friends
algorithms \cite{einasto1984,davis1985}. We used a friends-of-friends
halo
finder\footnote{http://www-hpcc.astro.washington.edu/tools/fof.html}
with the standard linking length equal to 20\% of the inter-particle
separation, $L_{\rm{box}}/\sqrt[3]{N_{\rm{part}}}$. In principle, this
linking length should be changed to take into account how
virialisation is different for different cosmologies and gravity
theories, however it is not clear how this should be done for
parametrised modified gravity and we have made the standard choice in
all of our calculations. It is now known that a systematic bias arises
when determining the mass of halos with the friends-of-friends
algorithm \cite{warren06}. As shown in \cite{warren06} this can be
corrected for by modifying the number of particles in each halo
according to the prescription $N_{\rm corr}=N\times(1-N^{-0.6})$,
where $N$ is the original number of clusters found by the
friends-of-friends algorithm in each mass interval and $N_{\rm corr}$
is the unbiased value. We have applied this correction to the halos
found in our simulations and also imposed a minimum number of
(uncorrected) particles per halo of 40. The halos were then binned
into logarithmic mass bins of width 0.2, between $10^{13}h^{-1}
M_{\odot}$ and $10^{15}h^{-1} M_{\odot}$ where $M_\odot$ represent a
solar mass.

\begin{figure}[t]
\begin{center}
\includegraphics[trim = 0mm 50mm 0mm 30mm, clip, width=3.2in]{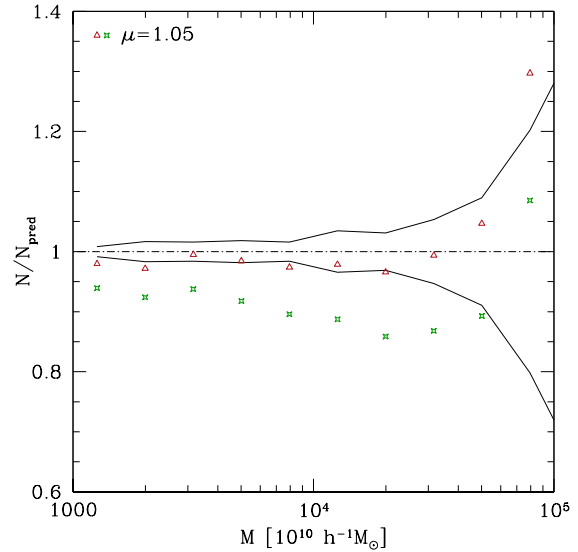}
\end{center}
\caption{The ratio of cluster counts found in the modified gravity simulations
  with respect to the mass function prediction. Both counts are for
  the same set of simulations that included a screening scale of 1
  Mpc. The green (crosses) points do not include the effect of
  the screening cut in the linear matter power spectrum used in the
  mass function prediction whilst the red (triangles) do. The result
  shows that the mass function formalism gives an accurate prediction
  of the counts if the screening scale is taken into account in the
  prediction.}
\label{fig:screen}
\end{figure}

\section{Results}\label{sec:res}

In this section we examine the output of the simulations at a
$z=0$. Firstly, we examine the change to cluster counts due to the
modification of gravity. Figure \ref{fig:total} shows the total number
of clusters found in the GR simulation, as well as the modified
gravity cases with and without a screening scale. We plot the average
value of the 10 independent realisations. The error bars show the
standard deviation estimated from the 10 samples. As expected, the
number of clusters increases due to the gravity being stronger in our
fiducial modified model with $\mu > 1$. The increase is larger towards
the higher masses. The screening scale reduces the increase in clusters in
modified gravity as we would naively expect. In addition, there is a
slight decrease in the number of cluster at the lowest masses. It is
not clear if this decrement is significant or physically relevant but
it may be consistent with the fact that at some mass scale the effect
of screening will mean that objects that would have formed clusters in
the GR case end up bound together with larger clusters which grow
faster on scales beyond the screening scale.

In figure \ref{fig:massfunc} we compare the numbers of clusters in our
simulations to the predictions in each case.  The linear matter power
spectra used as inputs to the W05 mass function are computed using the
MGCAMB code, with $\mu=1$ for the GR case, and $\mu=1.05$ for the
modified gravity case. For the modified gravity simulations with and
without the screening scale, the linear mass function was calculated
without the screening scale, as would be the case for models of
modified gravity where the screening scale is an entirely non-linear
phenomenon. The mass function works as well for the unscreened
modified gravity case as it does for GR. However, for the simulations
that incorporate a screening scale, the mass function prediction no
longer works. By inserting the screening scale into calculation of the
linear matter power spectrum we can reconcile the mass function
formalism with number counts observed in the simulations. The effect
of taking the screening scale into account in the linear power
spectrum used as input into the mass function is shown in figure
\ref{fig:screen}.

\section{Discussion}\label{sec:disc}

We have shown that the mass function formalism can work for simple
phenomenological models of modified gravity without a screening
scale. To model modified gravity with a screening scale, its effect
needs to be included in the calculation of the linear matter power
spectrum in order to retain the accuracy of the mass function
formalism. Of course, the mass function formalism is likely to break
down for more extreme departures from GR, however it isn't obvious
that large deviations are allowed by current data. In addition, more
complicated modelling of the $\mu$ parameter, such as including scale
dependence may reduce the accuracy of the mass function
predictions. However, as long as the scale dependence manifests in the
linear matter power spectrum and the deviations from GR are relatively
small then the mass function prediction may still be accurate. This
will require further work. The work presented here needs to be
extended in other ways as well, to more precisely determine the range
of validity of the mass function formalism. Varying the cosmological
parameters, linking length and parametrisation of $\mu$ will further
test the validity of the mass functions. In addition, running bigger
boxes and boxes with improved resolution will test the mass function
over a larger range of cluster masses. The mass function should also
be tested over a range of redshifts and with baryons included, rather
than solely dark matter particles. In addition, an examination of
spherical collapse in model independent modified gravity and/or an
examination of virialisation in these theories may shed further light
on finding halos and how mass functions work in modified gravity
simulations.

\bibliography{mn}

\begin{thebibliography}{36}
\expandafter\ifx\csname natexlab\endcsname\relax\def\natexlab#1{#1}\fi
\expandafter\ifx\csname bibnamefont\endcsname\relax
  \def\bibnamefont#1{#1}\fi
\expandafter\ifx\csname bibfnamefont\endcsname\relax
  \def\bibfnamefont#1{#1}\fi
\expandafter\ifx\csname citenamefont\endcsname\relax
  \def\citenamefont#1{#1}\fi
\expandafter\ifx\csname url\endcsname\relax
  \def\url#1{\texttt{#1}}\fi
\expandafter\ifx\csname urlprefix\endcsname\relax\def\urlprefix{URL }\fi
\providecommand{\bibinfo}[2]{#2}
\providecommand{\eprint}[2][]{\url{#2}}

\bibitem[{\citenamefont{{Thomas} and {Contaldi}}(2011)}]{paper2}
\bibinfo{author}{\bibfnamefont{D.~B.} \bibnamefont{{Thomas}}} \bibnamefont{and}
  \bibinfo{author}{\bibfnamefont{C.~R.} \bibnamefont{{Contaldi}}},
  \bibinfo{journal}{ArXiv e-prints}  (\bibinfo{year}{2011}),
  \eprint{1107.0727}.

\bibitem[{\citenamefont{{Clifton} et~al.}(2011)\citenamefont{{Clifton},
  {Ferreira}, {Padilla}, and {Skordis}}}]{mgreview}
\bibinfo{author}{\bibfnamefont{T.}~\bibnamefont{{Clifton}}},
  \bibinfo{author}{\bibfnamefont{P.~G.} \bibnamefont{{Ferreira}}},
  \bibinfo{author}{\bibfnamefont{A.}~\bibnamefont{{Padilla}}},
  \bibnamefont{and}
  \bibinfo{author}{\bibfnamefont{C.}~\bibnamefont{{Skordis}}},
  \bibinfo{journal}{ArXiv e-prints}  (\bibinfo{year}{2011}),
  \eprint{1106.2476}.

\bibitem[{\citenamefont{{Press} and {Schechter}}(1974)}]{ps}
\bibinfo{author}{\bibfnamefont{W.~H.} \bibnamefont{{Press}}} \bibnamefont{and}
  \bibinfo{author}{\bibfnamefont{P.}~\bibnamefont{{Schechter}}},
  \bibinfo{journal}{\apj} \textbf{\bibinfo{volume}{187}}, \bibinfo{pages}{425}
  (\bibinfo{year}{1974}).

\bibitem[{\citenamefont{{Bond} et~al.}(1991)\citenamefont{{Bond}, {Cole},
  {Efstathiou}, and {Kaiser}}}]{bond91}
\bibinfo{author}{\bibfnamefont{J.~R.} \bibnamefont{{Bond}}},
  \bibinfo{author}{\bibfnamefont{S.}~\bibnamefont{{Cole}}},
  \bibinfo{author}{\bibfnamefont{G.}~\bibnamefont{{Efstathiou}}},
  \bibnamefont{and} \bibinfo{author}{\bibfnamefont{N.}~\bibnamefont{{Kaiser}}},
  \bibinfo{journal}{\apj} \textbf{\bibinfo{volume}{379}}, \bibinfo{pages}{440}
  (\bibinfo{year}{1991}).

\bibitem[{\citenamefont{{Sheth} et~al.}(2001)\citenamefont{{Sheth}, {Mo}, and
  {Tormen}}}]{st99a}
\bibinfo{author}{\bibfnamefont{R.~K.} \bibnamefont{{Sheth}}},
  \bibinfo{author}{\bibfnamefont{H.~J.} \bibnamefont{{Mo}}}, \bibnamefont{and}
  \bibinfo{author}{\bibfnamefont{G.}~\bibnamefont{{Tormen}}},
  \bibinfo{journal}{\mnras} \textbf{\bibinfo{volume}{323}}, \bibinfo{pages}{1}
  (\bibinfo{year}{2001}), \eprint{arXiv:astro-ph/9907024}.

\bibitem[{\citenamefont{{Sheth} and {Tormen}}(1999)}]{st99b}
\bibinfo{author}{\bibfnamefont{R.~K.} \bibnamefont{{Sheth}}} \bibnamefont{and}
  \bibinfo{author}{\bibfnamefont{G.}~\bibnamefont{{Tormen}}},
  \bibinfo{journal}{\mnras} \textbf{\bibinfo{volume}{308}},
  \bibinfo{pages}{119} (\bibinfo{year}{1999}), \eprint{arXiv:astro-ph/9901122}.

\bibitem[{\citenamefont{{Sheth} and {Tormen}}(2002)}]{st02}
\bibinfo{author}{\bibfnamefont{R.~K.} \bibnamefont{{Sheth}}} \bibnamefont{and}
  \bibinfo{author}{\bibfnamefont{G.}~\bibnamefont{{Tormen}}},
  \bibinfo{journal}{\mnras} \textbf{\bibinfo{volume}{329}}, \bibinfo{pages}{61}
  (\bibinfo{year}{2002}), \eprint{arXiv:astro-ph/0105113}.

\bibitem[{\citenamefont{{Wang} and {Steinhardt}}(1998)}]{wang98}
\bibinfo{author}{\bibfnamefont{L.}~\bibnamefont{{Wang}}} \bibnamefont{and}
  \bibinfo{author}{\bibfnamefont{P.~J.} \bibnamefont{{Steinhardt}}},
  \bibinfo{journal}{\apj} \textbf{\bibinfo{volume}{508}}, \bibinfo{pages}{483}
  (\bibinfo{year}{1998}), \eprint{arXiv:astro-ph/9804015}.

\bibitem[{\citenamefont{{Battye} and {Weller}}(2003)}]{battye03}
\bibinfo{author}{\bibfnamefont{R.~A.} \bibnamefont{{Battye}}} \bibnamefont{and}
  \bibinfo{author}{\bibfnamefont{J.}~\bibnamefont{{Weller}}},
  \bibinfo{journal}{\prd} \textbf{\bibinfo{volume}{68}},
  \bibinfo{pages}{083506} (\bibinfo{year}{2003}),
  \eprint{arXiv:astro-ph/0305568}.

\bibitem[{\citenamefont{{Rapetti} et~al.}(2010)\citenamefont{{Rapetti},
  {Allen}, {Mantz}, and {Ebeling}}}]{rapetti10}
\bibinfo{author}{\bibfnamefont{D.}~\bibnamefont{{Rapetti}}},
  \bibinfo{author}{\bibfnamefont{S.~W.} \bibnamefont{{Allen}}},
  \bibinfo{author}{\bibfnamefont{A.}~\bibnamefont{{Mantz}}}, \bibnamefont{and}
  \bibinfo{author}{\bibfnamefont{H.}~\bibnamefont{{Ebeling}}},
  \bibinfo{journal}{\mnras} \textbf{\bibinfo{volume}{406}},
  \bibinfo{pages}{1796} (\bibinfo{year}{2010}), \eprint{0911.1787}.

\bibitem[{\citenamefont{{Shapiro} et~al.}(2010)\citenamefont{{Shapiro},
  {Dodelson}, {Hoyle}, {Samushia}, and {Flaugher}}}]{shapiro10}
\bibinfo{author}{\bibfnamefont{C.}~\bibnamefont{{Shapiro}}},
  \bibinfo{author}{\bibfnamefont{S.}~\bibnamefont{{Dodelson}}},
  \bibinfo{author}{\bibfnamefont{B.}~\bibnamefont{{Hoyle}}},
  \bibinfo{author}{\bibfnamefont{L.}~\bibnamefont{{Samushia}}},
  \bibnamefont{and}
  \bibinfo{author}{\bibfnamefont{B.}~\bibnamefont{{Flaugher}}},
  \bibinfo{journal}{\prd} \textbf{\bibinfo{volume}{82}},
  \bibinfo{pages}{043520} (\bibinfo{year}{2010}), \eprint{1004.4810}.

\bibitem[{\citenamefont{{Bhattacharya}
  et~al.}(2010)\citenamefont{{Bhattacharya}, {Heitmann}, {White}, {Luki{\'c}},
  {Wagner}, and {Habib}}}]{massf10}
\bibinfo{author}{\bibfnamefont{S.}~\bibnamefont{{Bhattacharya}}},
  \bibinfo{author}{\bibfnamefont{K.}~\bibnamefont{{Heitmann}}},
  \bibinfo{author}{\bibfnamefont{M.}~\bibnamefont{{White}}},
  \bibinfo{author}{\bibfnamefont{Z.}~\bibnamefont{{Luki{\'c}}}},
  \bibinfo{author}{\bibfnamefont{C.}~\bibnamefont{{Wagner}}}, \bibnamefont{and}
  \bibinfo{author}{\bibfnamefont{S.}~\bibnamefont{{Habib}}},
  \bibinfo{journal}{ArXiv e-prints}  (\bibinfo{year}{2010}),
  \eprint{1005.2239}.

\bibitem[{\citenamefont{{Courtin} et~al.}(2011)\citenamefont{{Courtin},
  {Rasera}, {Alimi}, {Corasaniti}, {Boucher}, and {F{\"u}zfa}}}]{courtin10}
\bibinfo{author}{\bibfnamefont{J.}~\bibnamefont{{Courtin}}},
  \bibinfo{author}{\bibfnamefont{Y.}~\bibnamefont{{Rasera}}},
  \bibinfo{author}{\bibfnamefont{J.-M.} \bibnamefont{{Alimi}}},
  \bibinfo{author}{\bibfnamefont{P.-S.} \bibnamefont{{Corasaniti}}},
  \bibinfo{author}{\bibfnamefont{V.}~\bibnamefont{{Boucher}}},
  \bibnamefont{and}
  \bibinfo{author}{\bibfnamefont{A.}~\bibnamefont{{F{\"u}zfa}}},
  \bibinfo{journal}{\mnras} \textbf{\bibinfo{volume}{410}},
  \bibinfo{pages}{1911} (\bibinfo{year}{2011}), \eprint{1001.3425}.

\bibitem[{\citenamefont{{Warren} et~al.}(2006)\citenamefont{{Warren},
  {Abazajian}, {Holz}, and {Teodoro}}}]{warren06}
\bibinfo{author}{\bibfnamefont{M.~S.} \bibnamefont{{Warren}}},
  \bibinfo{author}{\bibfnamefont{K.}~\bibnamefont{{Abazajian}}},
  \bibinfo{author}{\bibfnamefont{D.~E.} \bibnamefont{{Holz}}},
  \bibnamefont{and}
  \bibinfo{author}{\bibfnamefont{L.}~\bibnamefont{{Teodoro}}},
  \bibinfo{journal}{\apj} \textbf{\bibinfo{volume}{646}}, \bibinfo{pages}{881}
  (\bibinfo{year}{2006}), \eprint{arXiv:astro-ph/0506395}.

\bibitem[{\citenamefont{{Schmidt}
  et~al.}(2009{\natexlab{a}})\citenamefont{{Schmidt}, {Vikhlinin}, and
  {Hu}}}]{schmidt09c}
\bibinfo{author}{\bibfnamefont{F.}~\bibnamefont{{Schmidt}}},
  \bibinfo{author}{\bibfnamefont{A.}~\bibnamefont{{Vikhlinin}}},
  \bibnamefont{and} \bibinfo{author}{\bibfnamefont{W.}~\bibnamefont{{Hu}}},
  \bibinfo{journal}{\prd} \textbf{\bibinfo{volume}{80}},
  \bibinfo{pages}{083505} (\bibinfo{year}{2009}{\natexlab{a}}),
  \eprint{0908.2457}.

\bibitem[{\citenamefont{{Schmidt}
  et~al.}(2009{\natexlab{b}})\citenamefont{{Schmidt}, {Lima}, {Oyaizu}, and
  {Hu}}}]{schmidtfr}
\bibinfo{author}{\bibfnamefont{F.}~\bibnamefont{{Schmidt}}},
  \bibinfo{author}{\bibfnamefont{M.}~\bibnamefont{{Lima}}},
  \bibinfo{author}{\bibfnamefont{H.}~\bibnamefont{{Oyaizu}}}, \bibnamefont{and}
  \bibinfo{author}{\bibfnamefont{W.}~\bibnamefont{{Hu}}},
  \bibinfo{journal}{\prd} \textbf{\bibinfo{volume}{79}},
  \bibinfo{pages}{083518} (\bibinfo{year}{2009}{\natexlab{b}}),
  \eprint{0812.0545}.

\bibitem[{\citenamefont{{Schmidt}}(2009)}]{schmidt09b}
\bibinfo{author}{\bibfnamefont{F.}~\bibnamefont{{Schmidt}}},
  \bibinfo{journal}{\prd} \textbf{\bibinfo{volume}{80}},
  \bibinfo{pages}{123003} (\bibinfo{year}{2009}), \eprint{0910.0235}.

\bibitem[{\citenamefont{{Chan} and {Scoccimarro}}(2009)}]{chan09}
\bibinfo{author}{\bibfnamefont{K.~C.} \bibnamefont{{Chan}}} \bibnamefont{and}
  \bibinfo{author}{\bibfnamefont{R.}~\bibnamefont{{Scoccimarro}}},
  \bibinfo{journal}{\prd} \textbf{\bibinfo{volume}{80}},
  \bibinfo{pages}{104005} (\bibinfo{year}{2009}), \eprint{0906.4548}.

\bibitem[{\citenamefont{{Zhao} et~al.}(2010)\citenamefont{{Zhao}, {Macci{\`o}},
  {Li}, {Hoekstra}, and {Feix}}}]{zhao10}
\bibinfo{author}{\bibfnamefont{H.}~\bibnamefont{{Zhao}}},
  \bibinfo{author}{\bibfnamefont{A.~V.} \bibnamefont{{Macci{\`o}}}},
  \bibinfo{author}{\bibfnamefont{B.}~\bibnamefont{{Li}}},
  \bibinfo{author}{\bibfnamefont{H.}~\bibnamefont{{Hoekstra}}},
  \bibnamefont{and} \bibinfo{author}{\bibfnamefont{M.}~\bibnamefont{{Feix}}},
  \bibinfo{journal}{\apjl} \textbf{\bibinfo{volume}{712}},
  \bibinfo{pages}{L179} (\bibinfo{year}{2010}), \eprint{0910.3207}.

\bibitem[{\citenamefont{{Li} and {Barrow}}(2011)}]{li2010}
\bibinfo{author}{\bibfnamefont{B.}~\bibnamefont{{Li}}} \bibnamefont{and}
  \bibinfo{author}{\bibfnamefont{J.~D.} \bibnamefont{{Barrow}}},
  \bibinfo{journal}{\prd} \textbf{\bibinfo{volume}{83}},
  \bibinfo{pages}{024007} (\bibinfo{year}{2011}), \eprint{1005.4231}.

\bibitem[{\citenamefont{{Daniel} et~al.}(2010)\citenamefont{{Daniel}, {Linder},
  {Smith}, {Caldwell}, {Cooray}, {Leauthaud}, and {Lombriser}}}]{daniel10}
\bibinfo{author}{\bibfnamefont{S.~F.} \bibnamefont{{Daniel}}},
  \bibinfo{author}{\bibfnamefont{E.~V.} \bibnamefont{{Linder}}},
  \bibinfo{author}{\bibfnamefont{T.~L.} \bibnamefont{{Smith}}},
  \bibinfo{author}{\bibfnamefont{R.~R.} \bibnamefont{{Caldwell}}},
  \bibinfo{author}{\bibfnamefont{A.}~\bibnamefont{{Cooray}}},
  \bibinfo{author}{\bibfnamefont{A.}~\bibnamefont{{Leauthaud}}},
  \bibnamefont{and}
  \bibinfo{author}{\bibfnamefont{L.}~\bibnamefont{{Lombriser}}},
  \bibinfo{journal}{\prd} \textbf{\bibinfo{volume}{81}},
  \bibinfo{pages}{123508} (\bibinfo{year}{2010}), \eprint{1002.1962}.

\bibitem[{\citenamefont{{Pogosian} et~al.}(2010)\citenamefont{{Pogosian},
  {Silvestri}, {Koyama}, and {Zhao}}}]{gbz10b}
\bibinfo{author}{\bibfnamefont{L.}~\bibnamefont{{Pogosian}}},
  \bibinfo{author}{\bibfnamefont{A.}~\bibnamefont{{Silvestri}}},
  \bibinfo{author}{\bibfnamefont{K.}~\bibnamefont{{Koyama}}}, \bibnamefont{and}
  \bibinfo{author}{\bibfnamefont{G.}~\bibnamefont{{Zhao}}},
  \bibinfo{journal}{\prd} \textbf{\bibinfo{volume}{81}},
  \bibinfo{pages}{104023} (\bibinfo{year}{2010}), \eprint{1002.2382}.

\bibitem[{\citenamefont{{Khoury} and {Weltman}}(2004)}]{chameleon}
\bibinfo{author}{\bibfnamefont{J.}~\bibnamefont{{Khoury}}} \bibnamefont{and}
  \bibinfo{author}{\bibfnamefont{A.}~\bibnamefont{{Weltman}}},
  \bibinfo{journal}{\prd} \textbf{\bibinfo{volume}{69}},
  \bibinfo{pages}{044026} (\bibinfo{year}{2004}),
  \eprint{arXiv:astro-ph/0309411}.

\bibitem[{\citenamefont{{Vainshtein}}(1972)}]{vainshtein}
\bibinfo{author}{\bibfnamefont{A.~I.} \bibnamefont{{Vainshtein}}},
  \bibinfo{journal}{Physics Letters B} \textbf{\bibinfo{volume}{39}},
  \bibinfo{pages}{393} (\bibinfo{year}{1972}).

\bibitem[{\citenamefont{{Springel}}(2005)}]{gadget2}
\bibinfo{author}{\bibfnamefont{V.}~\bibnamefont{{Springel}}},
  \bibinfo{journal}{\mnras} \textbf{\bibinfo{volume}{364}},
  \bibinfo{pages}{1105} (\bibinfo{year}{2005}),
  \eprint{arXiv:astro-ph/0505010}.

\bibitem[{\citenamefont{{Bagla}}(2002)}]{treepm}
\bibinfo{author}{\bibfnamefont{J.~S.} \bibnamefont{{Bagla}}},
  \bibinfo{journal}{Journal of Astrophysics and Astronomy}
  \textbf{\bibinfo{volume}{23}}, \bibinfo{pages}{185} (\bibinfo{year}{2002}),
  \eprint{arXiv:astro-ph/9911025}.

\bibitem[{\citenamefont{{Lewis} et~al.}(2000)\citenamefont{{Lewis},
  {Challinor}, and {Lasenby}}}]{camb}
\bibinfo{author}{\bibfnamefont{A.}~\bibnamefont{{Lewis}}},
  \bibinfo{author}{\bibfnamefont{A.}~\bibnamefont{{Challinor}}},
  \bibnamefont{and}
  \bibinfo{author}{\bibfnamefont{A.}~\bibnamefont{{Lasenby}}},
  \bibinfo{journal}{\apj} \textbf{\bibinfo{volume}{538}}, \bibinfo{pages}{473}
  (\bibinfo{year}{2000}), \eprint{arXiv:astro-ph/9911177}.

\bibitem[{\citenamefont{{Zhao} et~al.}(2009)\citenamefont{{Zhao}, {Pogosian},
  {Silvestri}, and {Zylberberg}}}]{mgcamb}
\bibinfo{author}{\bibfnamefont{G.}~\bibnamefont{{Zhao}}},
  \bibinfo{author}{\bibfnamefont{L.}~\bibnamefont{{Pogosian}}},
  \bibinfo{author}{\bibfnamefont{A.}~\bibnamefont{{Silvestri}}},
  \bibnamefont{and}
  \bibinfo{author}{\bibfnamefont{J.}~\bibnamefont{{Zylberberg}}},
  \bibinfo{journal}{\prd} \textbf{\bibinfo{volume}{79}},
  \bibinfo{pages}{083513} (\bibinfo{year}{2009}), \eprint{0809.3791}.

\bibitem[{\citenamefont{{Jarosik} et~al.}(2011)\citenamefont{{Jarosik},
  {Bennett}, {Dunkley}, {Gold}, {Greason}, {Halpern}, {Hill}, {Hinshaw},
  {Kogut}, {Komatsu} et~al.}}]{wmap}
\bibinfo{author}{\bibfnamefont{N.}~\bibnamefont{{Jarosik}}},
  \bibinfo{author}{\bibfnamefont{C.~L.} \bibnamefont{{Bennett}}},
  \bibinfo{author}{\bibfnamefont{J.}~\bibnamefont{{Dunkley}}},
  \bibinfo{author}{\bibfnamefont{B.}~\bibnamefont{{Gold}}},
  \bibinfo{author}{\bibfnamefont{M.~R.} \bibnamefont{{Greason}}},
  \bibinfo{author}{\bibfnamefont{M.}~\bibnamefont{{Halpern}}},
  \bibinfo{author}{\bibfnamefont{R.~S.} \bibnamefont{{Hill}}},
  \bibinfo{author}{\bibfnamefont{G.}~\bibnamefont{{Hinshaw}}},
  \bibinfo{author}{\bibfnamefont{A.}~\bibnamefont{{Kogut}}},
  \bibinfo{author}{\bibfnamefont{E.}~\bibnamefont{{Komatsu}}},
  \bibnamefont{et~al.}, \bibinfo{journal}{\apjs}
  \textbf{\bibinfo{volume}{192}}, \bibinfo{pages}{14} (\bibinfo{year}{2011}),
  \eprint{1001.4744}.

\bibitem[{\citenamefont{{Scoccimarro}}(1998)}]{2lpta}
\bibinfo{author}{\bibfnamefont{R.}~\bibnamefont{{Scoccimarro}}},
  \bibinfo{journal}{\mnras} \textbf{\bibinfo{volume}{299}},
  \bibinfo{pages}{1097} (\bibinfo{year}{1998}),
  \eprint{arXiv:astro-ph/9711187}.

\bibitem[{\citenamefont{{Crocce} et~al.}(2006)\citenamefont{{Crocce},
  {Pueblas}, and {Scoccimarro}}}]{2lptb}
\bibinfo{author}{\bibfnamefont{M.}~\bibnamefont{{Crocce}}},
  \bibinfo{author}{\bibfnamefont{S.}~\bibnamefont{{Pueblas}}},
  \bibnamefont{and}
  \bibinfo{author}{\bibfnamefont{R.}~\bibnamefont{{Scoccimarro}}},
  \bibinfo{journal}{\mnras} \textbf{\bibinfo{volume}{373}},
  \bibinfo{pages}{369} (\bibinfo{year}{2006}), \eprint{arXiv:astro-ph/0606505}.

\bibitem[{\citenamefont{{Jenkins} et~al.}(2001)\citenamefont{{Jenkins},
  {Frenk}, {White}, {Colberg}, {Cole}, {Evrard}, {Couchman}, and
  {Yoshida}}}]{jenkins01}
\bibinfo{author}{\bibfnamefont{A.}~\bibnamefont{{Jenkins}}},
  \bibinfo{author}{\bibfnamefont{C.~S.} \bibnamefont{{Frenk}}},
  \bibinfo{author}{\bibfnamefont{S.~D.~M.} \bibnamefont{{White}}},
  \bibinfo{author}{\bibfnamefont{J.~M.} \bibnamefont{{Colberg}}},
  \bibinfo{author}{\bibfnamefont{S.}~\bibnamefont{{Cole}}},
  \bibinfo{author}{\bibfnamefont{A.~E.} \bibnamefont{{Evrard}}},
  \bibinfo{author}{\bibfnamefont{H.~M.~P.} \bibnamefont{{Couchman}}},
  \bibnamefont{and}
  \bibinfo{author}{\bibfnamefont{N.}~\bibnamefont{{Yoshida}}},
  \bibinfo{journal}{\mnras} \textbf{\bibinfo{volume}{321}},
  \bibinfo{pages}{372} (\bibinfo{year}{2001}), \eprint{arXiv:astro-ph/0005260}.

\bibitem[{\citenamefont{{Heitmann} et~al.}(2006)\citenamefont{{Heitmann},
  {Luki{\'c}}, {Habib}, and {Ricker}}}]{heit06}
\bibinfo{author}{\bibfnamefont{K.}~\bibnamefont{{Heitmann}}},
  \bibinfo{author}{\bibfnamefont{Z.}~\bibnamefont{{Luki{\'c}}}},
  \bibinfo{author}{\bibfnamefont{S.}~\bibnamefont{{Habib}}}, \bibnamefont{and}
  \bibinfo{author}{\bibfnamefont{P.~M.} \bibnamefont{{Ricker}}},
  \bibinfo{journal}{\apjl} \textbf{\bibinfo{volume}{642}}, \bibinfo{pages}{L85}
  (\bibinfo{year}{2006}), \eprint{arXiv:astro-ph/0601233}.

\bibitem[{\citenamefont{{Lacey} and {Cole}}(1994)}]{lacey1994}
\bibinfo{author}{\bibfnamefont{C.}~\bibnamefont{{Lacey}}} \bibnamefont{and}
  \bibinfo{author}{\bibfnamefont{S.}~\bibnamefont{{Cole}}},
  \bibinfo{journal}{\mnras} \textbf{\bibinfo{volume}{271}},
  \bibinfo{pages}{676} (\bibinfo{year}{1994}), \eprint{arXiv:astro-ph/9402069}.

\bibitem[{\citenamefont{{Einasto} et~al.}(1984)\citenamefont{{Einasto},
  {Klypin}, {Saar}, and {Shandarin}}}]{einasto1984}
\bibinfo{author}{\bibfnamefont{J.}~\bibnamefont{{Einasto}}},
  \bibinfo{author}{\bibfnamefont{A.~A.} \bibnamefont{{Klypin}}},
  \bibinfo{author}{\bibfnamefont{E.}~\bibnamefont{{Saar}}}, \bibnamefont{and}
  \bibinfo{author}{\bibfnamefont{S.~F.} \bibnamefont{{Shandarin}}},
  \bibinfo{journal}{\mnras} \textbf{\bibinfo{volume}{206}},
  \bibinfo{pages}{529} (\bibinfo{year}{1984}).

\bibitem[{\citenamefont{{Davis} et~al.}(1985)\citenamefont{{Davis},
  {Efstathiou}, {Frenk}, and {White}}}]{davis1985}
\bibinfo{author}{\bibfnamefont{M.}~\bibnamefont{{Davis}}},
  \bibinfo{author}{\bibfnamefont{G.}~\bibnamefont{{Efstathiou}}},
  \bibinfo{author}{\bibfnamefont{C.~S.} \bibnamefont{{Frenk}}},
  \bibnamefont{and} \bibinfo{author}{\bibfnamefont{S.~D.~M.}
  \bibnamefont{{White}}}, \bibinfo{journal}{\apj}
  \textbf{\bibinfo{volume}{292}}, \bibinfo{pages}{371} (\bibinfo{year}{1985}).

\end{thebibliography}
\end{document}